\documentclass[aip,rsi,reprint,amsmath,amssymb,author-year,author-numerical,floatfix]{revtex4-1}
\usepackage{graphicx}
\usepackage{dcolumn}
\usepackage{bm}
\usepackage[mathlines]{lineno}
\usepackage[utf8]{inputenc}
\usepackage[T1]{fontenc}
\usepackage{mathptmx}

\begin{document}
\title{A cryogenic-helium pipe flow facility with unique double-line molecular tagging velocimetry capability}

\author{H. Sanavandi}
\affiliation{Department of Mechanical Engineering, Florida State University, Tallahassee, Florida 32310, USA}
\affiliation{National High Magnetic Field Laboratory (NHMFL), Florida State University, 1800 E Paul Dirac Dr., Tallahassee, Florida 32310, USA}

\author{S. R. Bao}
\affiliation{Department of Mechanical Engineering, Florida State University, Tallahassee, Florida 32310, USA}
\affiliation{National High Magnetic Field Laboratory (NHMFL), Florida State University, 1800 E Paul Dirac Dr., Tallahassee, Florida 32310, USA}

\author{Y. Zhang}
\affiliation{Department of Mechanical Engineering, Florida State University, Tallahassee, Florida 32310, USA}
\affiliation{Florida Center for Advanced Aero-Propulsion (FCAAP), Florida State University, 2003 Levy Ave., Tallahassee, Florida 32310, USA}

\author{R. Keijzer}
\affiliation{National High Magnetic Field Laboratory (NHMFL), Florida State University, 1800 E Paul Dirac Dr., Tallahassee, Florida 32310, USA}

\author{W. Guo}
\email [Corresponding: ]{wguo@magnet.fsu.edu}
\affiliation{Department of Mechanical Engineering, Florida State University, Tallahassee, Florida 32310, USA}
\affiliation{National High Magnetic Field Laboratory (NHMFL), Florida State University, 1800 E Paul Dirac Dr., Tallahassee, Florida 32310, USA}

\author{L. N. Cattafesta III}
\affiliation{Department of Mechanical Engineering, Florida State University, Tallahassee, Florida 32310, USA}
\affiliation{Florida Center for Advanced Aero-Propulsion (FCAAP), Florida State University, 2003 Levy Ave., Tallahassee, Florida 32310, USA}

\begin{abstract}
Cryogenic helium-4 has extremely small kinetic viscosity, which makes it a promising material for high Reynolds ($Re$) number turbulence research in compact laboratory apparatuses. In its superfluid phase (He II), helium has an extraordinary heat transfer capability and has been utilized in various scientific and engineering applications. In order to unlock the full potential of helium in turbulence research and to improve our understanding of the heat transfer mechanism in He II, a flow facility that allows quantitative study of helium heat-and-mass transfer processes is needed. Here we report our work in assembling and testing a unique helium pipe flow facility that incorporates a novel double-line molecular tracking velocimetry (DL-MTV) system. This flow facility allows us to generate turbulent pipe flows with $Re$ above $10^7$, and it can also be adapted to produce heat-induced counterflow in He II. The DL-MTV system, which is based on the generation and tracking of two parallel thin He$^*_2$ molecular tracer lines with an adjustable separation distance, allows us to measure not only the velocity profile but also both the transverse and longitudinal spatial velocity structure functions. We have also installed a deferential pressure sensor to the flow pipe for pressure drop measurement. The testing results of the flow facility and the measurement devices are presented. We discuss how this facility will allow us to solve some outstanding problems in the helium heat-and-mass transfer topic area.
\end{abstract}
\maketitle

\section{Introduction}\label{SecI}
Cryogenic helium-4 ($^4$He) is known for its great potential in fluid mechanics research and in thermal engineering applications due to its unique mechanical and thermal properties \cite{Sreenivasan-2001-AAM}. For instance, the kinematic viscosity $\nu$ of liquid $^4$He can be lower than 10$^{-8}$ m$^2$/s, which is about three orders of magnitudes smaller than that for ambient air \cite{Donnelly-1998-JPCRD}. Therefore, it is feasible to generate turbulent flows in liquid helium with an extremely high Reynolds ($Re$) number (defined as $Re=DU/\nu$, where $D$ and $U$ represent the characteristic length and velocity of the flow). Understanding such high $Re$ flows can benefit the design of transportation vehicles and defense vessels for better control and improved energy efficiency. Compared to existing high $Re$ flow facilities that utilize more conventional fluid materials, a cryogenic flow facility using liquid $^4$He has some unique advantages. For instance, the State-of-the-art Princeton Superpipe facility uses compressed air up to 220 bar to achieve the desired low kinematic viscosity \cite{Zagarola-1997-PRL}. This high pressure makes it very challenging to incorporate view ports in the flow facility for visualization measurement of the velocity field. On the other hand, quantitative flow visualization of liquid helium flows in compact cryostat has been demonstrated \cite{Guo-2014-PNAS}. Especially, a powerful molecular tagging velocimetry (MTV) technique has been developed in our lab \cite{Gao-2015-RSI}, which allows us to measure both the instantaneous velocity profile of the $^4$He flow in a channel and the spatial velocity structure functions \cite{Marakov-2015-PRB, Gao-2016-PRB, Gao-2016-JETP, Gao-2017-PRB, Varga-2018-PRB, Gao-2018-PRB, Bao-2018-PRB}. Such measurements are largely impractical for conventional high $Re$ flow facilities that rely on single-point flow measurement tools. Nonetheless, our MTV technique has not yet been implemented and demonstrated in any helium-based high $Re$ flow equipment.

Besides its small kinematic viscosity, helium is also known for its fascinating quantum hydrodynamics in the superfluid phase. Below about 2.17 K, ordinary liquid $^4$He (He I) transits to the superfluid phase (He II), which consists of two fully miscible components: an inviscid superfluid component with density $\rho_s$ (i.e., the condensate) and a viscous normal-fluid component with density $\rho_n$ (i.e., the thermal excitations) \cite{Tilley-book}. This two-fluid system has many interesting properties. For instance, instead of ordinary convective, heat transfer in He II is via an extremely effective counterflow mode: the normal fluid carries the heat away from the heat source, and the superfluid, which carries no entropy, flows in the opposite direction to compensate the fluid mass. Another unique feature of He II is that the rotational motion in the superfluid can occur only with the formation of topological defects in the form of quantized vortex lines \cite{Donnelly-1991-book}. These vortex lines all have identical cores (about 1~{\AA} in radius) and they each carry a single quantum of circulation $\kappa\simeq10^{-3}$ cm/s. Turbulence in the superfluid therefore takes the form of an irregular tangle of vortex lines (quantum turbulence) \cite{Vinen-2002-JLTP}. The normal fluid behaves more like a classical fluid. But a force of mutual friction between the two fluids \cite{Vinen-1957-PRS}, arising from the scattering of thermal excitations by the vortex lines, can affect the flows in both fluids. This mutual friction can significantly alter the turbulence characteristics as well as the boundary-layer profile of He II in various flows. Studying novel emergent behaviors of this two-fluid system often requires a flow facility with the capability of measuring both the longitudinal and transverse spatial velocity structure functions, which is lacking at present.

\begin{figure*}[htb]
\includegraphics [width=0.95 \hsize]{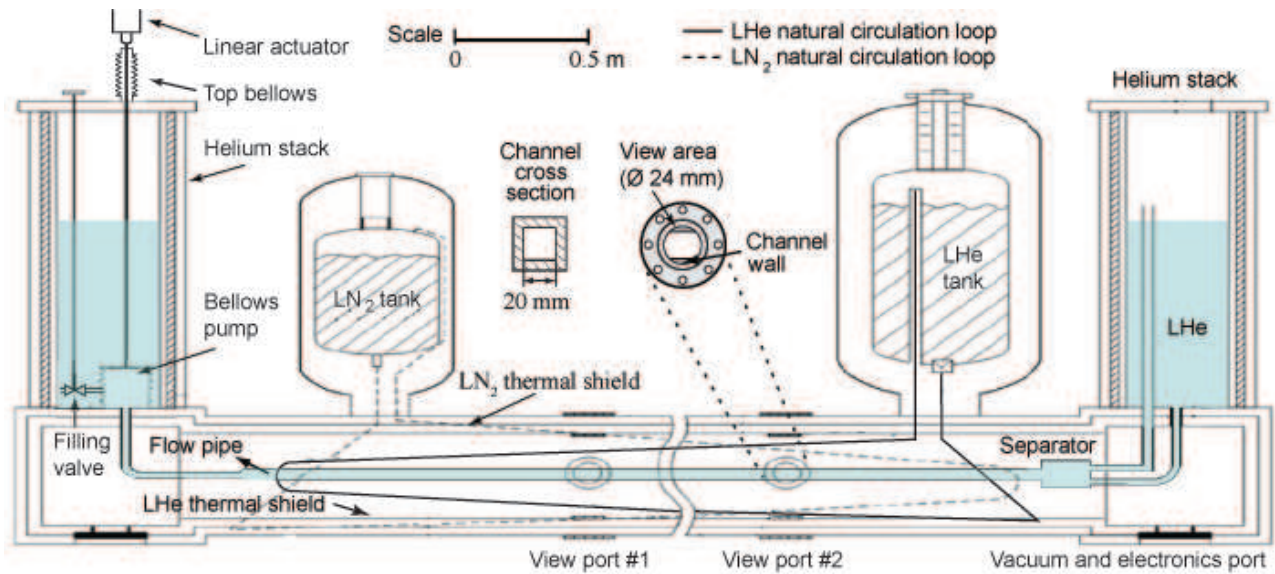}
\caption{A schematic diagram of the Liquid Helium Flow Visualization Facility (LHFVF).}
\label{Fig1}
\end{figure*}

In this paper, we discuss our work on assembling and testing a unique helium pipe flow facility that incorporates a novel double-line molecular tracking velocimetry (DL-MTV) measurement system. This flow facility is adapted from an existing liquid helium flow visualization facility (LHFVF) that was built and utilized by Van Sciver and colleagues \cite{Xu-2007-PF,Chagovets-2015-PF}. Turbulent pipe flows in liquid $^4$He with $Re$ above $10^7$ can be produced, and it can also be adapted to produce thermal counterflow in He II. The DL-MTV system, which is upgraded based on our existing MTV optics, allows the generation and tracking of two parallel thin He$^*_2$ molecular tracer lines in liquid helium with an adjustable separation distance. With this DL-MTV system, the near-wall velocity profile and the velocity structure functions in both the longitudinal and transverse directions can be obtained. A deferential pressure sensor is also incorporated in the LHFVF for pressure drop measurement across the flow pipe. In Sec~\ref{SecII}, we describe the experimental apparatus which include the adapted LHFVF, the laser and imaging systems, and the pressure sensor. The testing results of these apparatus are presented in Sec~\ref{SecIII}. In Sec.~\ref{SecIV}, we discuss how this facility will allow us to solve some outstanding problems in the helium heat-and-mass transfer topic area. A brief summary is given in Sec.~\ref{SecV}.

\section{Experimental Apparatus}\label{SecII}
\subsection{Liquid Helium Flow Visualization Facility}
The Liquid Helium Flow Visualization Facility (LHFVF) is an cryostat designed for generating and visualizing liquid $^4$He pipe flows. This facility consists of a horizontal cylindrical experimental space (5 m long with an inner diameter of 0.2 m) surrounded by two concentric radiation shields that are cooled by natural convection loops from the liquid helium and liquid nitrogen tanks (see the schematic diagram in Fig.~\ref{Fig1}). These shields and the tanks all sit inside the evacuated cryostat body. A flow pipe with a square cross-section (2$\times$2 cm$^2$) and a length of 3.35 m is installed at the center of the experimental space. This pipe is connected to two vertical helium storage stacks at the two ends of the LHFVF. The temperature of the helium in the stacks can be controlled by regulating the vapor pressure. For flow visualization purpose, the LHFVF is equipped with two sets of view ports, one at the midpoint and one about 1 m downstream. Each window set consists of aligned windows (top, bottom, and front) installed in the flow pipe, the radiation shields, and the cryostat body. The windows mounted on the helium shield are coated with infrared reflective film to minimize the radiation heat leak to the experimental space. The front window in the flow pipe has a diameter of 24 mm, greater than the inner side-width of pipe (see the inset in Fig.~\ref{Fig1}). This design allows us to examine the boundary layer flow in the vicinity of the pipe wall. In the experiment, we pass the laser beams through the top and bottom windows and place the camera near the front window for image acquisition.

To generate the flow in the pipe, in the original LHFVF setup \cite{Xu-2007-PF,Chagovets-2015-PF}, two bellows pumps were installed (one in each helium stack) and were welded to the flow pipe. These bellows pumps were supposed to move always oppositely such that the liquid $^4$He can be pushed to flow through the pipe from one bellows to the other. However, during the last operation a few years ago these bellows pumps were severely damaged due to a malfunction of the control unit in coordinating the motions of the two bellows. This facility was since put in storage until we restored it recently. In the current LHFVF setup, we removed the broken bellows and installed a single bellows pump in the left stack (see Fig.~\ref{Fig1}). A superfluid leak-tight cryogenic filling valve is welded to the bellows, which controls liquid $^4$He feeding into the bellows. The new bellows has an effective cross-section area 1.81$\times 10^{-2}$ m$^2$ and a stroke length of 9.4 cm, which provides a maximum volume displacement of about 1.7 liter of liquid $^4$He. The bellows is connected through a rod to a linear actuator (Parker ETS32) mounted coaxially on the top of the left stack. A computer-controlled stepper motor (Parker S57-102) is used to drive the linear actuator. This stepper motor has a limiting thrust of 600 N, which is more than enough to drive the low-viscosity liquid $^4$He through the pipe even at the highest speed we have tested.

\subsection{Double-line Molecular Tagging System}
\begin{figure}[!tb]
\includegraphics [width=1 \hsize]{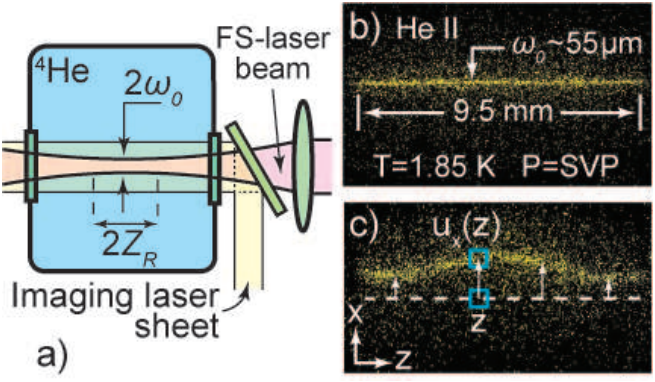}
\caption{(a) A schematic diagram showing the typical experimental setup for creating and imaging He$_2^*$ molecular tracer lines in helium. (b) A representative tracer-line image in He II upon its creation. (c) A schematic showing how the local velocity is calculated based on the displacement of line segments.}
\label{Fig2}
\end{figure}
In order to make quantitative velocity measurement of the $^4$He flows in the LHFVF, we have implemented the MTV technique that was developed in our lab \cite{Gao-2015-RSI}. The tracer particles used in the MTV are He$_2^*$ molecules in the excited electron-spin triplet state. These excimer molecules can be easily created in helium as a consequence of ionization or excitation of the ground state $^4$He atoms \cite{Dennis-1969-PRL,Hill-1971-PRL}. They form tiny bubbles in liquid $^4$He (about 6~{\AA} in radius \cite{Benderskii-2002-JCP}) and have an exceptionally long lifetime (about 13 s~\cite{McKinsey-PRA-1999}). Due to their small size and effective mass, He$_2^*$ molecules always follow the fluid motion in gaseous helium and He I, and they are entrained by the viscous normal-fluid component in He II since the Stokes drag easily dominates other forces \cite{McKinsey-PRL-2005}.

In our previous MTV experiments, a 5-kHz femtosecond (fs) laser system (wavelength $\lambda$: 780 nm, duration: 35 fs, pulse energy: up to 4 mJ) was used to generate thin lines of He$_2^*$ tracers in helium via laser-field ionization \cite{Gao-2015-RSI}. As shown schematically in Fig.~\ref{Fig2} (a), the fs-laser beam is focused by a lens with a focal length $f$ and is passed through an optical cryostat that contains helium at a regulated pressure and temperature. For an ideal Gaussian beam with a beam radius $\omega_0$ at the focal plane, one can define a Rayleigh range $z_R=\pi{\omega_0}^2/\lambda$, over which the laser intensity drops by 50\% due to beam spreading \cite{Self-1983-AO}. The He$^*_2$ tracers are expected to be produced essentially within the Rayleigh range. Our past tests showed that a fs-laser pulse energy of about 60 $\mu$J is sufficient to create He$^*_2$ tracers. We then send in 3-5 pulses from a 1-kHz imaging laser at 905 nm to drive the He$_2^*$ tracers to produce 640 nm fluorescent light \cite{Guo-PRL-2009, Guo-JLTP-2010, Guo-PRL-2010}. The fluorescence is captured by an intensified CCD (ICCD) camera mounted perpendicular to the tracer-line plane. Fig.~\ref{Fig1} (b) shows a typical fluorescence image of the He$^*_2$ tracer line taken right after its creation in He II. The width of the tracer line is about $2\omega_0$ and its length is about $2Z_R$ as expected. To extract velocity information, we allow an initially straight tracer line to move with the fluid by a drift time $\triangle t$ (see Fig.~\ref{Fig1} (c)). The deformed tracer line is divided into small segments and the center of each segment can be determined by a Gaussian fit of its intensity profile. When $\triangle t$ is small, the streamwise velocity $u_x(z)$ can be calculated as the displacement of the segment at $z$ divided by $\triangle t$ \cite{Miles-ARFM-1997}. This MTV method has been successfully applied to study various types of turbulent flows in He II \cite{Marakov-2015-PRB, Gao-2016-PRB, Gao-2016-JETP, Gao-2017-PRB, Varga-2018-PRB, Gao-2018-PRB, Bao-2018-PRB}.

\begin{figure}[htb]
\includegraphics [width=0.75 \hsize]{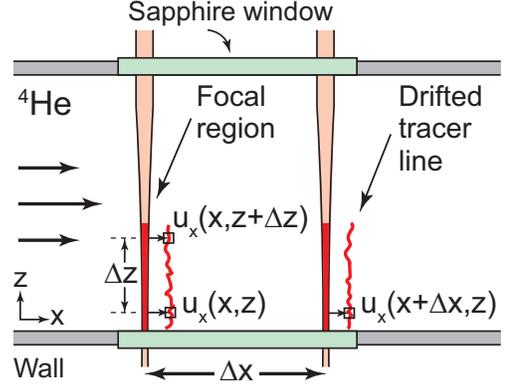}
\caption{A schematic diagram showing the concept of the double-line molecular tagging velocimetry (DL-MTV).}
\label{Fig3}
\end{figure}
Despite the usefulness of the MTV technique, by tracking a single tracer line, we can only correlate the measured streamwise velocities along the tracer line to determine the $n$-th order transverse velocity structure function $S_n^{\perp }(\vec{r})$ but cannot get any information about the longitudinal velocity structure function $S_n^{\parallel }(\vec{r})$. These structure functions are defined as:
\begin{eqnarray}
\begin{split}
S_n^{\perp }(\vec{r})=\overline{|u_x (x,z+\Delta z)-u_x (x,z)|^n} \\
S_n^{\parallel }(\vec{r})=\overline{|u_x (x+\Delta x,z)-u_x (x,z)|^n}
\end{split}
\end{eqnarray}
where $x$ and $z$ are, respectively, the coordinates in the streamwise and the transverse directions, and the overline denotes ensemble averaging. Knowing these structure functions, one can extract quantitative information about the energy spectrum and other statistical properties of the turbulent flows \cite{Davidson-2004-book}. In the case that the flows to be examined are boundary flows or anisotropic turbulent flows where the scalings of these structure functions are very different, it is highly desirable to have the capability of measuring both of them \cite{Dhruva-1997-PRE,Grossmann-1997-PF}. To achieve this goal, a feasible solution is to create two parallel tracer lines in the flow pipe with an adjustable separation distance, as shown schematically in Fig.~\ref{Fig3}. By tracking the displacement of the two tracer lines, one can determine the streamwise velocities along both lines. Then, by correlating the velocities at two locations in the streamwise direction and in the transverse direction, both $S_n^{\parallel }(\vec{r})$ and $S_n^{\perp }(\vec{r})$ can be obtained.

\begin{figure*}[htb]
\includegraphics [width=0.88 \hsize]{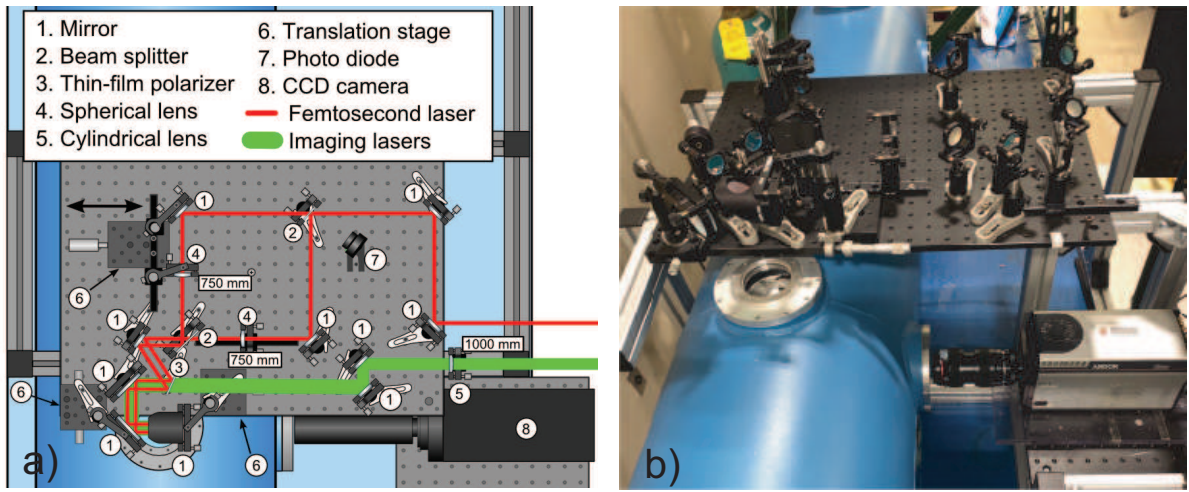}
\caption{(a) A schematic diagram showing the optical setup for creating two tracer lines. (b) A picture of the assembled optical components.}
\label{Fig4}
\end{figure*}

To implement this double-line MTV (DL-MTV) scheme, we have designed and assembled a unique optical system. First, a periscope device (i.e., a vertical post with two mirror-sets installed at its two ends) is used to guide both the fs-laser and the imaging laser beams from the optical table to a breadboard installed on top of the LHFVF. Then, the fs-laser beam is divided into two orthogonal beams in parallel with the breadboard using a beam splitter, as shown schematically in Fig.~\ref{Fig4} (a). One of the two beams is reflected on a mirror that is mounted on a translation stage such that the separation distance between the two fs-beams can be continuously adjusted from zero to a maximum separation of about 10 mm with a sub-micron resolution. The 905-nm imaging laser is focused by a cylindrical lens into a laser sheet (thickness: 1 mm, width: 10 mm) that covers the entire region traversed by the two tracer lines. Finally, the two fs-beams are focused by two separate spherical lenses before they are combined with the imaging laser sheet using a polarizer-based beam combiner and reflected vertically down through the LHFVF. Note that the two spherical lenses are mounted on optical rails such that they can be easily moved without affecting the parallelity or orientation of the fs-laser beams. The advantage of these movable lenses is that we can then control the creation of the two tracer lines at arbitrary distances from the bottom wall of the flow pipe. This feature is especially useful for examining the near-wall velocity profile. Fig.~\ref{Fig4} (b) shows a picture of the optical components that we have assembled. 

\subsection{Pressure Sensor}
\begin{figure}[h!]
\includegraphics [width=1 \hsize]{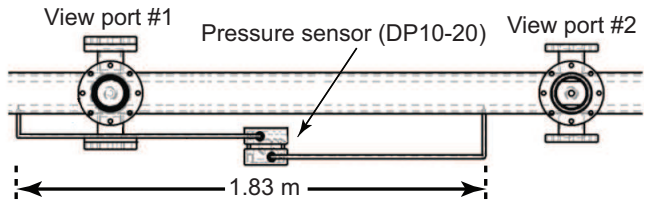}
\caption{A schematic diagram showing the locations where the pressure drop is measured by the pressure sensor.}
\label{Fig5}
\end{figure}

In pipe flow research and applications, a useful parameter for evaluating the frictional lose is the friction factor $f_D$. For an impressible fluid, this factor is related to the pressure drop $\Delta P$ along the flow pipe as $\Delta P=f_D\cdot(\frac{1}{2}\rho {\bar{u}}^2)\cdot L_f/D_h$, where $\rho$ is the fluid density, $\bar{u}$ is the mean velocity in the pipe, and $D_h$ and $L_f$ are the hydraulic diameter and the length of the pipe, respectively \cite{Landau-book}. To enable the measurement of $f_D$ for helium pipe flows, we have installed a Validyne DP10-20 variable-reluctance deferential pressure transducer (DPT) to the flow pipe inside the LHFVF, as shown schematically in Fig.~\ref{Fig5}.

This DPT sensor has a flat diaphragm sensing element clamped between cases halves that are connected through stainless-steel tubes (0.125-inch in diameter) to the bottom wall of the flow pipe at locations separated by $L_f=1.83$ m. Special silver-coated indium-brazed stainless-steel gaskets are used to reliably seal the sensor to the tubes to prevent leakage in the He II runs. A Validyne CD19A carrier demodulator module is utilized to excite the sensor and to read the voltage output. This voltage signal can be calibrated and converted to pressure readings. We choose DP10-20 because of its very good linearity in signal response. The nominal pressure-difference range of the DP10-20 sensor is $0-860$ Pa. Nevertheless, it is specified that the sensor's response can remain linear up to 200\% full pressure span with less that 0.5\% zero shift, which nicely covers the range of the anticipated pressure drop in our experiments. However, since these specifications are designated for operation temperatures above 220 K, the sensor performance needs to be tested in helium.

\section{Test Results}\label{SecIII}
\subsection{Flow Facility Testing}
We have developed a very effective procedure to cool down the big LHFVF. First, the cryostat vacuum space is evacuated to below 10$^{-3}$ Pa. The nitrogen and the helium radiation shields are then cooled by introducing cryogenic liquids into the respective tanks. After that, cold helium vapor from a liquid helium storage dewar is forced to flow from the right stack through the flow pipe to the left stack with the filling valve open. This procedure efficiently pre-cools both stacks and the flow pipe due to the maximal usage of the vapor enthalpy. When the temperature of the stacks and the flow pipe drop to below 15 K, we start transferring liquid $^4$He into the right stack. After both stacks are fully filled, we then pump on the stacks to cool the liquid helium to a desired temperature by regulating the pressures in the stacks. We have successfully cooled down the LHFVF and achieved a helium temperature as low as 1.4 K using this procedure.

\begin{figure}[htb]
\includegraphics [width=0.95 \hsize]{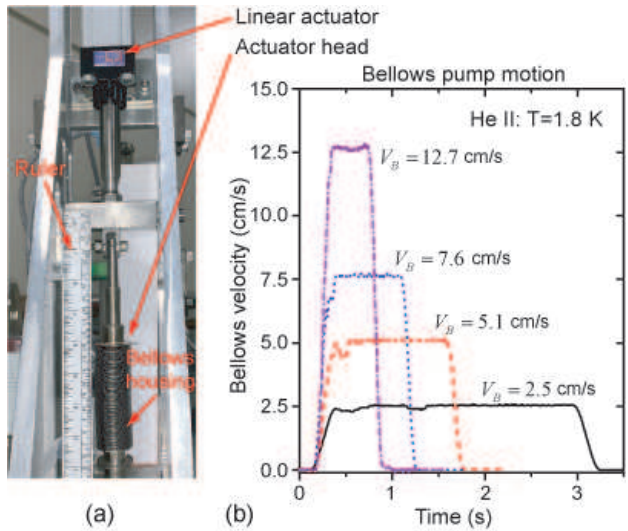}
\caption{(a) A picture showing the linear actuator, the bellows housing, and the ruler placed nearby for scale calibration. (b) Representative curves showing the measured bellows velocity versus time.}
\label{Fig6}
\end{figure}

To generate flows in the pipe, we close the filling valve and then push the bellows pump by controlling the stepper motor using a LabVIEW computer program. This program sends commands to an indexer (6200 Parker Automation) that controls three key parameters of the bellows motion: the steady bellows velocity $V_B$, its transient acceleration $a_B$, and the total displacement $\Delta h$ (which is always less than 9.4 cm). To test the actual performance of the bellows pump and the control system, we utilize a high-speed CCD camera (IDT XS-3) to take consecutive images of the linear actuator's head. A ruler is placed nearby ro provide length scale calibration, as shown in Fig.~\ref{Fig6} (a). The instantaneous velocity of the actuator (and hence the bellows) can be determined by analyzing the obtained images.

Test runs with $V_B$ in the range of 0.025 cm/s to 12.7 cm/s have been conducted while the stacks and the flow pipe were filled with He II at 1.8 K. The total displacement is set to $\Delta h=7.6$, and the acceleration up to $a_B=127$ cm/s$^2$ is used. Representative velocity curves are shown in Fig.~\ref{Fig6} (b). All cases show a good agreement between the actual steady velocity and the programmed velocity between the transient acceleration and deceleration regions. Note that the ratio of cross-section areas between the bellows pump and the flow pipe is about 45.3. Therefore, the bellows velocities that we have tested correspond to mean flow velocities of He II in the flow pipe as $\bar{u}\in$[0.01, 5.75] m/s. Using the known properties of He II \cite{Donnelly-1998-JPCRD}, one can work out that the pipe flow Reynolds number $Re_D$ is in the range of $2.2\times10^4$ to $1.3\times10^7$. Due to the finite stroke length, there is a limited time window $\Delta t_W$ for flow measurements (i.e., approximated $\Delta h/V_B$ when $a_B$ is high). Nevertheless, even at the highest velocity that we have tested, $\Delta t_W$ is still long enough for the development of the turbulent flow and for us to make quantitative velocity field and pressure drop measurements.

\subsection{DL-MTV System Testing}
Upon the completion of the DL-MTV optical setup, we have carefully tuned the entire laser and imaging system and conducted tests to ensure that the DL-MTV scheme is indeed achieved. These tests include the alignment and overlapping of the fs-laser and imaging laser beams, visual examination of their beam profiles, and Rayleigh-range measurement of the fs-beams for controlling the thicknesses and lengths of the tracer lines. After that, the optical setup is tested for producing and for position-control of the He$_2^*$ tracer lines in the flow pipe filled with liquid helium.

\begin{figure}[!tb]
	\includegraphics [width=0.85 \hsize]{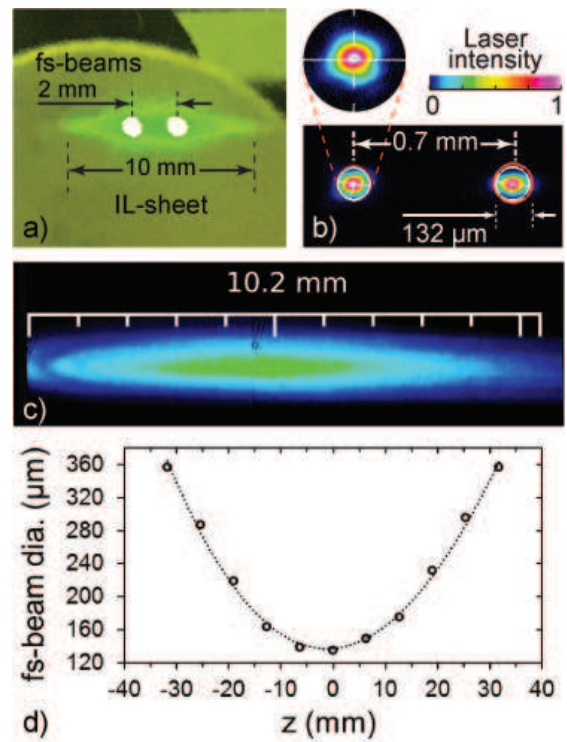}
	\caption{(a) A picture showing the overlapping cross sections of the fs-laser and imaging laser beams on an IR card. (b) and (c) are representative images showing the cross-section intensity profiles of the two fs-laser beams and the imaging laser sheet taken with the IR camera. (d) The measured diameter $2\omega(z)$ along one of the fs-laser beams. The solid curve represents a fit using the ideal Gaussian beam equation as discussed in the text.}
	\label{Fig7}
\end{figure}

To ensure that the fs-laser and imaging laser beams have the desired overlapping as they pass through the flow pipe, we fine tune their directions independently using mirror pairs on the breadboard. An infrared (IR) card is then utilized to examine the cross-section profiles and the relative positions of the laser beams at locations both above the top view port of the LHFVF and below the bottom view port. Beam collimation is achieved when the relative positions of the laser beams do not shift from the top view port to the bottom view port. A typical beam profile picture on the IR card is included in Fig.~\ref{Fig7} (a), which clearly shows that the two fs-laser beams are covered within the imaging laser sheet.

In order to obtain more quantitative beam profile information, we remove the mirror that reflects the combined beams down into the LHFVF so that an IR camera (WinCamD-LCM4 from DataRay Inc.) can be placed at the focal region of the laser beams for beam profile measurement. Typical images of the cross-section intensity profiles of the fs-laser beams and the imaging-laser beam are shown in Fig.~\ref{Fig7} (b) and (c), respectively. The two fs-laser beams have circular cross sections with nearly Gaussian intensity profiles. Their separation distance can be easily adjusted using the movable mirror on the breadboard in the range 0 to 10 mm with a sub-micron resolution. The imaging laser sheet has a thickness of about 1 mm (measured at half the maximum intensity) and a width of 10.2 mm, close to our design specifications. We have also performed the beam width measurement along the two fs-laser beams in order to determine their Rayleigh range in the focal region. Fig.~\ref{Fig7} (d) shows an representative curve of the measured fs-beam diameter 2$\omega(z)$ along one of the fs-laser beams that is focused by a lens with $f=75$ cm. The measured beam diameter variation can be well fitted using the equation for ideal Gaussian beams \cite{Self-1983-AO}: $\omega(z)=\omega_0[1+(z/z_R)^2]$. From this fit, both the beam waist at the focal plane $\omega_0$ and the Rayleigh range $z_R$ can be determined. In the specific case shown in Fig.~\ref{Fig7} (d), $\omega_0=66$ $\mu$m and $z_R=16$ mm. Since $2z_R=32$ mm is greater than the inner width of the flow pipe (20 mm), we expect that a tracer line with a diameter of about $2\omega_0$ will be created through the entire width of the pipe. Note that for a Gaussian beam, when the focal length $f$ of the lens is much greater than $z_R$, the beam waist $\omega_0$ is given by \cite{Self-1983-AO}: $\omega_0=\lambda f/\pi\omega_L$, where $\lambda$ is the fs-laser wavelength and $\omega_L$ is the incident beam radius on the lens. Therefore, by using lenses with different $f$, we can easily control the thickness and length of the tracer lines. Tracer lines with $\omega_0$ as small as 10-20 $\mu$m has been readily created \cite{Gao-2015-RSI}.

\begin{figure}[htb]
	\includegraphics [width=0.95 \hsize]{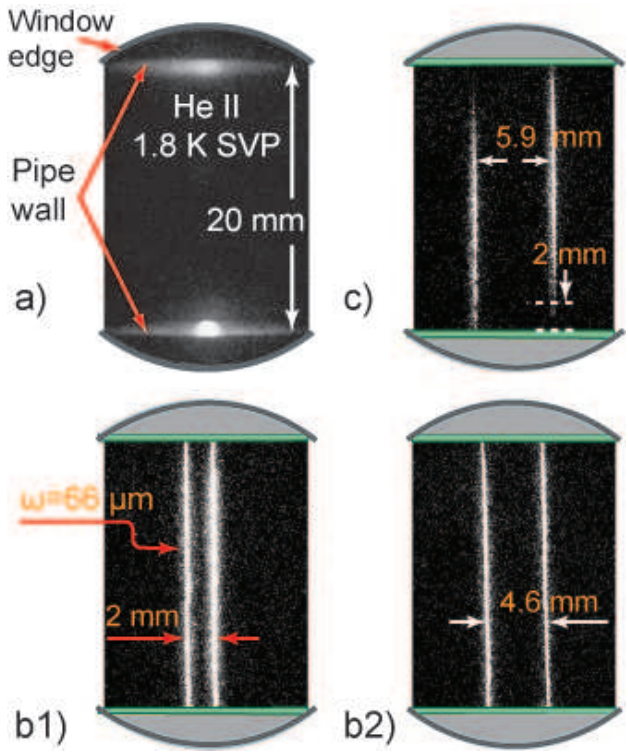}
	\caption{(a) A picture of the flow pipe illuminated with ambient light from the top view port. (b1) and (b2) are representative images showing the two tracer lines created at different streamwise separation distances using lenses with $f=75$ cm. c) A representative image demonstrating the tunability of the vertical positions of the two tracer lines created using lenses with $f=50$ cm.}
	\label{Fig8}
\end{figure}

Finally, we fill the flow pipe with liquid helium for tracer-line imaging test. Electronic shutters are used to allow 5 fs-laser pulses (at 5 kHz) to pass through the LHFVF to create the tracer lines which are then illuminated by a train of 3 imaging-laser pulses (at 1 kHz). Typical fluorescence images of two parallel tracer lines created in static He II at 1.8 K are shown in Fig. \ref{Fig8}. For reference, an image of the flow pipe taken with ambient light illumination from the top view port is included in Fig. \ref{Fig8} (a), where one can clearly see the top and bottom walls of the pipe. The streamwise separation distance between the two tracer lines can be easily adjusted, as demonstrated in Fig. \ref{Fig8} (b1) and (b1). We have also tested the tunability of the vertical positions of the tracer lines. In this case, lenses with $f=50$ cm are used so that the created tracer lines are shorter. Then, by adjusting the positions of the lenses in the breadboard, we can shift one line close to the bottom wall and one line close to the top wall of the pipe, as shown in Fig. \ref{Fig8} (c). This tunability is important when we create very thin tracer lines for measuring the near-wall velocity field.

\subsection{Pressure Sensor Testing}
Since the nominal specifications of the DP10-20 pressure sensor are not applicable for liquid helium temperatures, we have performed calibration of the sensor immersed in liquid helium in a test cryostat at a controlled bath temperature in the range 1.5--4.2 K. The sensing element of the pressure sensor is connected through pipes to helium gas reservoirs at controlled pressures. This way, the change in the voltage reading $\Delta V$ can be correlated with the actual pressure difference $\Delta P$ across the sensing element. Representative calibration data in He II at 1.8 K are shown in Fig. \ref{Fig8}. The sensor response remains linear to a maximum pressure difference of about 2500 Pa. Through a linear fit to the data, the conversion factor between the voltage change $\Delta V$ and the pressure drop $\Delta P$ can be determined. It turns out that this conversion factor only varies by a few percent from room temperature down to the lowest temperature we have tested.
\begin{figure}[htb]
	\includegraphics [width=1.0 \hsize]{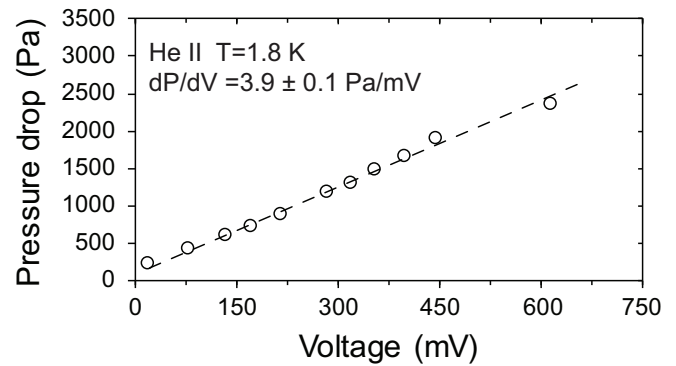}
	\caption{Representative calibration data of the DP10-20 pressure sensor immersed in He II.}
	\label{Fig9}
\end{figure}

\begin{figure}[!tb]
	\includegraphics [width=0.95 \hsize]{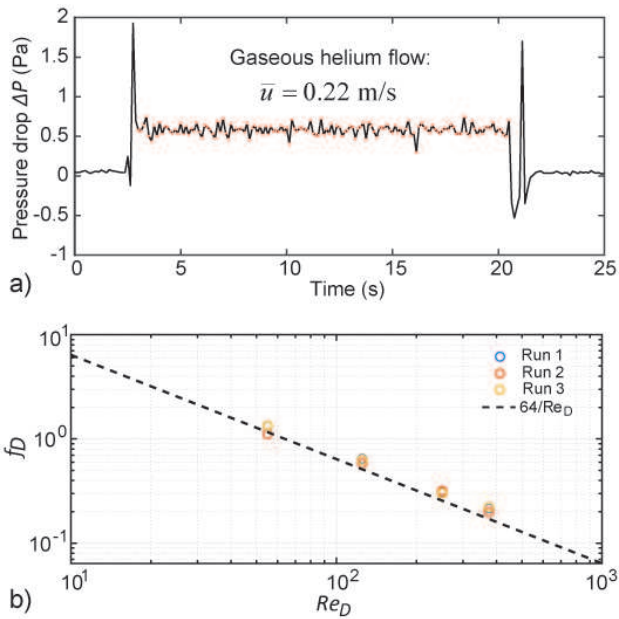}
	\caption{(a) A representative pressure-drop curve when a gaseous helium flow is generated in the pipe. (b) The calculated friction factor $f_D$ versus the Reynolds number $Re_D$ for the gas flows.}
	\label{Fig10}
\end{figure}

The calibrated pressure sensor is then installed to the flow pipe, and a test with flowing gaseous helium at 225 K is conducted to demonstrate the sensitivity of the sensor. By controlling the speed of the bellows pump, we generate gas flows in the pipe at four different mean velocities: $u=$0.22, 0.5, 1.0, and 1.5 m/s. Due to the low density of the gas ($\rho=0.214$ kg/m$^3$) and its relatively large viscosity ($\mu=16.39$ Pa$\cdot$s), the Reynolds number $Re_D=\rho u D_h/\mu$ at these velocities are approximately 57, 130, 260 and 390. Fig.~\ref{Fig10} (a) shows the pressure drop reading at $u=$0.22 m/s. It is impressive to see that a pressure drop as small as about 0.5 Pa is clearly resolved. Based on the measured pressure drop data, we can calculate the corresponding friction factor as $f_D=\Delta P(D_h/L_f)/(\frac{1}{2}\rho {\bar{u}}^2)$. Fig.~\ref{Fig10} (b) shows the obtained $f_D$ as a function of the Reynolds number $Re_D$, which indeed agrees very well with the expected friction factor behavior $f_D=64/Re_D$ for laminar flows. This agreement confirms that the pressure sensor is functioning well.

\section{Discussions}\label{SecIV}

The flow facility that we have assembled and tested has a number of unique features: 1) it allows the generation of high $Re$ pipe flows in both the classical fluid He I and in the quantum fluid He II; 2) the DL-MTV capability makes it possible to measure not only the instantaneous velocity profile in the pipe but also the longitudinal and transverse velocity structure functions; and 3) the pressure drop (and hence the friction factor) of the pipe flow can be measured. The combination of all these features makes the facility one of its kind in classical and quantum fluids research. In what follows, we outline a few interesting topics that can be studied using our facility.

\textbf{Law of the wall in classical pipe flow:} Turbulent pipe flow is a topic of great practical importance. In the traditional view, the near-wall profile of the mean velocity $\bar{u}$ in pipe flow can be described by a logarithmic form known as the “law of the wall”: $\bar{u}/u_{\tau}=(1/\kappa)\ln(z/\eta)+B$, where $u_{\tau}$ is the friction velocity that can be evaluated based on the pressure drop measurement \cite{Zagarola-1998-JFM}, $\eta$ is the viscous scale, $z$ denotes the wall-normal coordinate, and $\kappa$ and $B$ are the von K\'arm\'an coefficient and the additive constant \cite{Smits-2011-ARFM}. Despite extensive experimental and numerical investigations, there are still unresolved fundamental issues such as the extent of such a log law, the value of the log-law constants, and their $Re$ dependence \cite{Marusic-2010-PF,McKeon-2007-PTRSA}. So far, the state-of-the-art Princeton Superpipe experiments have observed the log law at $z>600\eta$ when $Re$ is greater than 2.3$\times$10$^5$. \cite{Zagarola-1997-PRL,Hultmark-2012-PRL} However, their reported K\'arm\'an coefficient $\kappa=0.42$ differs from typical values (i.e. $0.37-0.39$) found in high $Re$ boundary layer flows and channel flows \cite{Marusic-2010-PF}, which casts doubt on the universality of $\kappa$. On the other hand, Furuichi \emph{et al.} reported $\kappa=0.385$ in their recent experiment using the “Hi-Redff” pipe flow facility \cite{Furuichi-2015-PF}, supporting the universality of $\kappa$. Since the accurate value of $\kappa$ is crucial to modeling wall-bounded flows, more high-$Re$ pipe-flow measurements using independent facilities like ours is needed.

To resolve the near-wall velocity profile, it is crucial to have a fine dimensionless spatial resolution $l^+$, defined as the ratio of the probe size $l$ to the viscous scale $\eta$: $l^+=l/\eta$. The $l$ of our DL-MTV system is limited by the minimum drift distance of the tracer lines that can be resolved, which is about the half thickness of the lines. As we have discussed in Sec.~\ref{SecIII}, by using appropriate lenses, it is possible to achieve 10-20 $\mu$m for $l$. The viscous scale $\eta$ can be estimated as $\eta\approx D\cdot Re^{-0.75}$, where $D$ denotes the energy containing scale that roughly equals the hydraulic diameter of the pipe \cite{Davidson-2004-book}. If we plug in $l\sim20$~$\mu$m and $Re\sim10^6$, a dimensionless resolution $l^+$ of about 30 is obtained, which is comparable to the typical $l^+$ values in the Superpipe experiments and should be sufficient to resolve the logarithmic velocity profile that is expected to appear at $z^+=z/\eta>$600. Furthermore, by correlating velocities along individual tracer lines or between the two lines, new knowledge about the transverse and longitudinal velocity structure functions can be obtained using our DL-MTV. These correlation measurements are useful for understanding the eddy structures in pipe flows that are not possible to study using conventional single-point probes.

\begin{figure}[h!]
\includegraphics [width=0.8 \hsize]{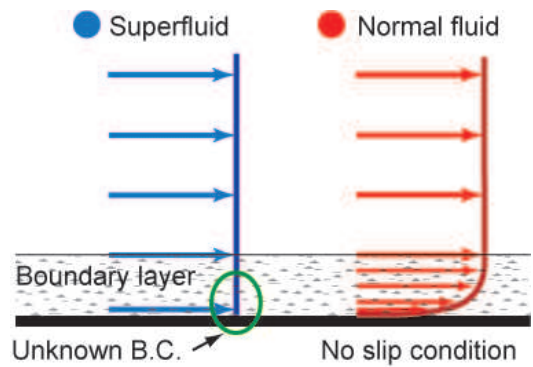}
\caption{A schematic showing possible mismatch of the velocity profiles of the two fluids in forced He II pipe flow.}
\label{Fig11}
\end{figure}

\textbf{Law of the wall in He II pipe flow:} Another interesting topic is the near-wall velocity profile in He II pipe flow. Many large-scale cooling systems for particle accelerators and superconducting magnets involve pipelines for transporting He II from the liquifiers or storage vessels to the equipment to be cooled \cite{VanSciver-book}. The friction factor $f_D$ of He II in pipe flow is needed in the design of these cooling systems. Despite some limited measurements of $f_D$ in He II \cite{Fuzier-2001-Cryo}, a clear understanding of its behavior is difficult without detailed knowledge of the near-wall velocity profile of the viscous normal fluid. This profile could deviate from the classical law-of-the-wall due to the mechanism as illustrated schematically in Fig.~\ref{Fig11}.

In He II pipe flow, the two fluids can become strongly coupled by mutual friction in the bulk liquid at scales greater than the mean vortex-line spacing \cite{Barenghi-1997-PF,Baggaley-2012-EPL}. However, the situation may change near the pipe wall. Due to the no-slip boundary condition of the viscous normal fluid, there is a strong velocity gradient in a very thin boundary layer. There is no guarantee that the mutual friction could be effective enough to maintain a similar velocity gradient in the superfluid. Such a velocity boundary layer in the super fluid would require highly nonuniform distribution and polarization of the quantized vortices pinned to the wall \cite{Stagg-2017-PRL}, about which there is no existing knowledge. Therefore, the two fluids could have mismatched velocity profiles near the pipe wall. The relative velocity $u_{ns}$ in the boundary layer then leads to a mutual friction $f_{ns}$ per unit volume between the two fluids that modifies the classical logarithmic velocity profile of the normal fluid. Measuring the revised law-of-the-wall of the normal fluid using our DL-MTV technique will not only enrich our knowledge of boundary-layer flows in general but also benefit various He II pipe-flow based applications.

\textbf{He II counterflow turbulence:}
In He II thermal counterflow, the velocity of the normal fluid $u_n$ is controlled by the heat flux $q$ as $q=\rho s T u_n$, where $\rho$ is the total density of He II and $s$ is the specific entropy \cite{Landau-book}. This heat transfer mode is extremely efficient and can lead to an effective thermal conductivity of He II higher than that for pure metals \cite{VanSciver-book}. Therefore, He II has been widely utilized for cooling scientific and industrial equipment such as superconducting magnets, power transmission cables, superconducting accelerator cavities, and satellites \cite{VanSciver-book}. However, it has been known that when the heat flux exceeds a small critical value, turbulence can appear spontaneously in the superfluid as a tangle of quantized vortices \cite{Vinen-1957-PRS}, which impairs the superior heat transfer capability of He II. In the past, most of the experimental and numerical studies have focused on the vortex-tangle dynamics in the superfluid \cite{Vinen-1957-PRS, Schwarz-1977-PRL, Barenghi-2001-book}. In recent years, by using our MTV technique, we have revealed that the normal fluid can also become turbulent and can exhibit non-classical scaling behaviors \cite{Marakov-2015-PRB, Gao-2016-PRB, Gao-2016-JETP, Gao-2017-PRB, Gao-2018-PRB, Bao-2018-PRB}. Indeed, due to the relative motion of the two fluids, the mutual friction sets in and dissipates the turbulent kinetic energy at all lengths scales, which is in marked contrast to classical turbulence where the energy dissipation is important only below the viscous scale $\eta$ \cite{Landau-book}. Understanding the novel normal-fluid turbulence and its influence on the vortex-tangle dynamics now becomes an outstanding challenging problem in quantum fluids research.

Interestingly, in a recent theoretical study, Biferale \emph{et al.} suggested that counterflow turbulence should exhibit strong anisotropy at small scales \cite{Biferale-2019-PRL}, which differs strongly from classical flows where better isotropy is expected at smaller scales. This intriguing property may be responsible for some unexplained behaviors of counterflow turbulence. The \emph{only} way to experimentally study this anisotropy effect is by using the DL-MTV that we have implemented in the LHFVF. Instead of pushing the bellows pump to generate pipe flows in the LHFVF, it is straightforward to install a heater inside the bellows to induce counterflow in the flow pipe. Then, by examining the displacement of two parallel tracer lines, we can compare the scaling behaviors of the longitudinal and the transverse velocity structure functions and thereby evaluate the anisotropy quantitatively. This study will greatly improve our understanding of the novel doubly turbulence in He II counterflow.

\section{Summary}\label{SecV}
We have assembled and tested a unique LHFVF that incorporates a novel DL-MTV measurement system. This flow facility allows the generation of turbulent pipe flows with $Re$ above $10^7$. The DL-MTV system allows us to measure not only the instantaneous velocity profile of He I (or the normal fluid in He II) but also the transverse and longitudinal spatial velocity structure functions. Besides, a pressure sensor has been installed and tested for pressure drop measurement. These measurement capabilities together with the various flows that can be generated make this facility an extremely useful equipment in classical and quantum fluids research. In particular, by studying the law of the wall in pipe flows of both He I and He II and by experimentally quantifying the novel anisotropy effect in He II thermal counterflow turbulence, new knowledge of cryogenic helium mass-and-heat transfer will be obtained, which will benefit various practical applications involving cryogenic helium.

\section* {Acknowledgements}
This work is supported by National Science Foundation under Grant No. CBET-1801780. The work was conducted at the National High Magnetic Field Laboratory, Florida State University, which is supported by the National Science Foundation Cooperative Agreement No. DMR-1644779 and the state of Florida.

\bibliographystyle{unsrt}

\bibliography{Reference}

\begin{thebibliography}{99}
\bibitem{Sreenivasan-2001-AAM} K. R. Sreenivasan and R. J. Donnelly, ``Role of cryogenic helium in classical fluid dynamics: Basic research and model testing'', Adv. Appl. Mech. \textbf{37}, 239 (2001).
\bibitem{Donnelly-1998-JPCRD} J. Donnelly and C. F. Barenghi, ``The Observed Properties of Liquid Helium at the Saturated Vapor Pressure'', J. Phys. Chem. Ref. Data \textbf{27}, 1217 (1998).
\bibitem{Zagarola-1997-PRL} M.V. Zagarola, A.J. Smits, ``Scaling of the Mean Velocity Profile for Turbulent Pipe Flow'', Phys. Rev. Lett., \textbf{78}, 239-242 (1997).
\bibitem{Guo-2014-PNAS} W. Guo, D. P. Lathrop, M. La Mantia, and S.W. Van Sciver, ``Visualization of two-fluid flows of superfluid helium-4 at finite temperatures'', Proc. Natl. Acad. Sci., \textbf{111}, 4653 (2014).
\bibitem{Gao-2015-RSI} J. Gao, A. Marakov, W. Guo, B.T. Pawlowski, S.W. Van Sciver, G.G. Ihas, D.N. McKinsey, and W.F. Vinen, ``Producing and Imaging a Thin Line of He$_2$ Tracer Molecules in Helium-4'', Rev. Sci. Instrum., \textbf{86}, 093904 (2015).
\bibitem{Marakov-2015-PRB} A. Marakov, J. Gao, W. Guo, S.W. Van Sciver, G.G. Ihas, D.N. McKinsey, and W.F. Vinen, ``Visualization of the normal-fluid turbulence in counterflowing superfluid $^4$He'', Phys. Rev. \textbf{B 91}, 094503 (2015).
\bibitem{Gao-2016-PRB} J. Gao, W. Guo, and W.F. Vinen, ``Determination of the effective kinematic viscosity for the decay of quasiclassical turbulence in superfluid $^4$He'', Phys. Rev. \textbf{B 94}, 094502 (2016).
\bibitem{Gao-2016-JETP} J. Gao, W. Guo, V.S. L'vov, A. Pomyalov, L. Skrbek, E. Varga, and W.F. Vinen, Challenging Problem in Quantum Turbulence: Decay of Counterflow in Superfluid $^4$He. JETP Letters, \textbf{103}, 732 (2016).
\bibitem{Gao-2017-PRB} J. Gao, E. Varga, W. Guo, and W.F. Vinen, ``Energy spectrum of thermal counterflow turbulence in superfluid helium-4'', Phys. Rev. \textbf{B 96}, 094511 (2017).
\bibitem{Varga-2018-PRB} E. Varga, J. Gao, W. Guo, and L. Skrbek, ``Intermittency enhancement in quantum turbulence in superfluid $^4$He'', Phys. Rev. Fluids, \textbf{3}, 094601 (2018).
\bibitem{Gao-2018-PRB} J. Gao, W. Guo, W.F. Vinen, S. Yui, and M. Tsubota, Dissipation in quantum turbulence in superfluid $^4$He. Phys. Rev. \textbf{B 97}, 184518 (2018).
 \bibitem{Bao-2018-PRB} S. Bao, W. Guo, V. S. L'vov, and Anna Pomyalov, “Statistics of turbulence and intermittency enhancement in superfluid $^4$He counterflow”, Phys. Rev. B 98, 174509 (2018).

\bibitem{Tilley-book} D.R. Tilley and J. Tilley, \emph{Superfluidity and superconductivity} (A. Hilger; University of Sussex Press, Boston, 1986), 2nd ed.
\bibitem{Donnelly-1991-book} R.J. Donnelly, \emph{Quantized vortices in helium II}, (Cambridge University Press, Cambridge England; New York, 1991).
\bibitem{Vinen-2002-JLTP} W.F. Vinen and J.J. Niemela, ``Quantum turbulence'', J. Low Temp. Phys., \textbf{129}, 213-213 (2002).
\bibitem{Vinen-1957-PRS} W.F. Vinen, ``Mutual Friction in a Heat Current in Liquid Helium II. I. Experiments on Steady Heat Currents'', Proc. Roy. Soc., \textbf{A 240}, 114-127 (1957).
\bibitem{Xu-2007-PF} T. Xu, and S. W. Van Sciver, ``Particle image velocimetry measurements of the velocity profile in He II forced flow'', Phys. Fluids, \textbf{19}, 071703 (2007).
\bibitem{Chagovets-2015-PF} T. V. Chagovets  and S. W. Van Sciver, ``Visualization of He II forced flow around a cylinder'', Phys. Fluids \textbf{27}, 045111 (2015).
\bibitem{Dennis-1969-PRL} W.S. Dennis, E. Durbin, W. Fitzsimm, O. Heybey, and G.K. Walters, ``Spectroscopic Identification of Excited Atomic and Molecular States in Electron-Bombarded Liquid Helium'', Phys. Rev. Lett. \textbf{23}, 1083 (1969).
\bibitem{Hill-1971-PRL} J.C. Hill, O. Heybey, and G.K. Walters, ``Evidence of Metastable Atomic and Molecular Bubble States in Electron-Bombarded Superfluid Liquid Helium'', Phys. Rev. Lett. \textbf{26}, 1213 (1971).
\bibitem{Benderskii-2002-JCP} A.V. Benderskii, J. Eloranta, R. Zadoyan, and V.A. Apkarian, ``A direct interrogation of superfluidity on molecular scales'', J. Chem. Phys., \textbf{117}, 1201-1213 (2002).
\bibitem{McKinsey-PRA-1999} D.N. McKinsey, C.R. Brome, J.S. Butterworth, S.N. Dzhosyuk, P.R. Huffman, C.E.H. Mattoni, J.M. Doyle, R. Golub, and K. Habicht, Radiative decay of the metastable He$_2$($a^3\Sigma^{+}_u$) molecule in liquid helium. Phys. Rev. \textbf{A 59}, 200-204 (1999).
\bibitem{McKinsey-PRL-2005} D.N. McKinsey, W.H. Lippincott, J.A. Nikkel, and W.G. Rellergert, Trace detection of metastable helium molecules in superfluid helium by laser-induced fluorescence. Phys. Rev. Lett., \textbf{95}, 111101 (2005).
\bibitem{Self-1983-AO} S.A. Self, ``Focusing of Spherical Gaussian Beams'', Appl. Optics, \textbf{22}, 658-661 (1983).
\bibitem{Guo-PRL-2009} W. Guo, J.D. Wright, S.B. Cahn, J.A. Nikkel, and D.N. McKinsey, Metastable Helium Molecules as Tracers in Superfluid He-4. Phys. Rev. Lett., \textbf{102}, 235301 (2009).
\bibitem{Guo-JLTP-2010} W. Guo, J.D. Wright, S.B. Cahn, J.A. Nikkel, and D.N. McKinsey, Studying the Normal-Fluid Flow in Helium-II Using Metastable Helium Molecules. J. Low Temp. Phys., \textbf{158}, 346-352 (2010).
\bibitem{Guo-PRL-2010} W. Guo, S.B. Cahn, J.A. Nikkel, W.F. Vinen, and D.N. McKinsey, Visualization study of counterflow in superfluid $^4$He using metastable helium molecules. Phys. Rev. Lett., \textbf{105}, 045301 (2010).
\bibitem{Miles-ARFM-1997} R.B. Miles and W.R. Lempert, Quantitative flow visualization in unseeded flows. Annu. Rev. Fluid Mech., \textbf{29}, 285 (1997)
\bibitem{Davidson-2004-book} P.A. Davidson, \emph{Turbulence: An Introduction for Scientist s and Engineers}, (Oxford University Press, Unite Kindom, 2004).
\bibitem{Dhruva-1997-PRE} B. Dhruva, Y. Tsuji, and K.R. Sreenivasan, Transverse structure functions in high-Reynolds-number turbulence, Phys. Rev. \textbf{E 56}, R4928 (1997).
\bibitem{Grossmann-1997-PF} S. Grossmann, D. Lohse, and A. Reeh, Different intermittency for longitudinal and transversal turbulent fluctuations, Phys. Fluids \textbf{9}, 3817 (1997).
\bibitem{Landau-book} L.D. Landau and E.M. Lifshitz, Fluid mechanics. (Pergamon Press, Oxford, England ; New York, 1987), 2nd ed.

\bibitem{Zagarola-1998-JFM} M.V. Zagarola and A.J. Smits, Mean-flow scaling of turbulent pipe flow, J. Fluid Mech., 373,  33 (1998).
\bibitem{Smits-2011-ARFM} A.J. Smits, B.J. McKeon, and I. Marusic, ``High-Reynolds Number Wall Turbulence'', Annu. Rev. Fluid Mech. \textbf{43}, 353 (2011).
\bibitem{Marusic-2010-PF} I. Marusic, B.J. McKeon, P.A. Monkewitz, H.M. Nagib, A.J. Smits, and K.R. Sreenivasan, ``Wall-bounded turbulent flows at high Reynolds numbers: Recent advances and key issues'', Phys. Fluids, \textbf{22}, 065103 (2010).
\bibitem{McKeon-2007-PTRSA} J. McKeon and K.R. Sreenivasan, ``Introduction: scaling and structure in high Reynolds number wall-bounded flows'', Phil. Trans. R. Soc. \textbf{A 365}, 635 (2007).
\bibitem{Hultmark-2012-PRL} M. Hultmark, M. Vallikivi, S.C.C. Bailey, and A.J. Smits, ``Turbulent Pipe Flow at Extreme Reynolds Numbers'', Phys. Rev. Lett. \textbf{108}, 094501 (2012).
\bibitem{Furuichi-2015-PF} N. Furuichi, Y. Terao, Y. Wada, and Y. Tsuji, ``Friction factor and mean velocity profile for pipe flow at high Reynolds numbers'', Phys. Fluids, \textbf{27}, 095108 (2015).
\bibitem{Fuzier-2001-Cryo} S. Fuzier, B. Baudouy, and S. W. Van Sciver, ``Steady-State Pressure Drop and Heat Transfer in He II Forced Flow at High Reynolds Number'', Cryogenics, \textbf{41}, 453-459 (2001).
\bibitem{Barenghi-1997-PF} C.F. Barenghi, D.C. Samuels, G.H. Bauer, and R.J. Donnelly, ``Superfluid vortex lines in a model of turbulent flow'', Phys. Fluids, \textbf{9}, 2631 (1997).
\bibitem{Baggaley-2012-EPL} A.W. Baggaley, C.F. Barenghi, A. Shukurov, and Y.A. Sergeev, ``Coherent vortex structures in quantum turbulence'', Europhys. Lett., \textbf{98}, 26002 (2012).
\bibitem{Stagg-2017-PRL} G.W. Stagg, N.G. Parker, and C.F. Barenghi, ``Superfluid Boundary Layer'', Phys. Rev. Lett. \textbf{118}, 135301 (2017).
\bibitem{VanSciver-book} S.W. Van Sciver, \emph{Helium Cryogenics}, (Springer, Boston, MA, United States, 2012).
\bibitem{Schwarz-1977-PRL} K. W. Schwarz, ``Theory of Turbulence in Superfluid He-4'', Phys. Rev. Lett. \textbf{38}, 551-554 (1977).
\bibitem{Barenghi-2001-book} C. F. Barenghi, R. J. Donnelly, and W. F. Vinen, \emph{Quantized vortex dynamics and superfluid turbulence}, (Springer Berlin Heidelberg, Germany, 2008).
\bibitem{Biferale-2019-PRL} L. Biferale, D. Khomenko, V. L'vov, A. Pomyalov, I. Procaccia, and G. Sahoo, ``Superfluid Helium in Three-Dimensional Counterflow Differs Strongly from Classical Flows: Anisotropy on Small Scales'', Phys. Rev. Lett. \textbf{122}, 144501 (2019).
\end{thebibliography}

\end{document}